# Anomalies in Light Scattering: A Circuit Model Approach

Deepanshu Trivedi[1], Arjuna Madanayake[1], and Alex Krasnok[1,2*]

[1]Department of Electrical and Computer Engineering, Florida International University, Miami, Florida 33174, USA

[2]Knight Foundation School of Computing and Information Sciences, Florida International University, Miami, FL 33199, USA

*To whom correspondence should be addressed: akrasnok@fiu.edu

**Abstract**

In experimental physics, it is essential to understand electromagnetic (EM) wave scattering across EM spectrum, from radio waves to X-rays, and is pivotal in driving photonics innovations. Recent advancements have uncovered phenomena like bound states in the continuum (BICs) and parity-time (PT) symmetric systems, which are closely associated with the characteristics of the scattering matrix and are governed by passivity and causality. The emergence of complex frequency excitations has transcended the constraints imposed by passivity and causality in a system, revealing effects such as virtual critical coupling and virtual gain. However, applying the concepts of complex frequency excitation in more complicated systems remains challenging. In this work, we demonstrate the extension of the lumped element model of circuit theory to the analysis of anomalies in light scattering in the complex frequency domain. We demonstrate that the circuit model approach can facilitate design and analysis of effects such as virtual perfect absorption, BICs, real and virtual critical coupling, exceptional points, and anisotropic transmission resonances (ATRs). These findings broaden comprehension of EM wave phenomena and pave the way for significant advancements in photonics, offering new methods for designing and optimizing optical devices and systems with broad-ranging applications.

**I. Introduction**

Electromagnetic wave scattering is a fundamental experimental technique used across the EM spectrum, from radio waves to X-rays. Understanding scattering theory is crucial for interpreting experimental data and comprehending physics. The manipulation of light scattering remains vital in photonics, driving discoveries and enhancing Nanophotonics in fields such as optical antennas [1], imaging [2], optical tweezers [3], and optical trapping [4]. Recent progress has unveiled numerous exceptional scattering phenomena in engineered structures, introducing effects such as bound states in the continuum (BIC) [5], [6], parity-time (PT) symmetry in non-Hermitian systems [7], real and virtual absorption, lasing, and nonradiating states [8]–[10]. These phenomena are explained by linking them to the poles and zeros of the scattering matrix in the complex frequency plane [8]. Implementing these effects necessitates design, a task made challenging by the complexity of the optical system. Moreover, passivity, causality, and energy conservation limit control over scattering, thus impacting the performance of optical devices.

A recent advancement in EM scattering is the use of complex frequency excitation to overcome traditional limitations imposed by energy conservation along with passivity and causality [8], [11]–[15]. This approach, involving exponentially growing or decaying signals, has revealed new phenomena including virtual critical coupling [12], [16], optical pulling forces [14], virtual parity-time symmetry [13], the non-Hermitian skin effect [17], transient PT-symmetry [18], and superlensing [19]. These discoveries represent a step forward, opening avenues in studying and applying EM wave scattering. While progress in complex frequency EM scattering has been notable, its applications have predominantly been confined to simple systems. This limitation stems from the analytical work needed to tailor the excitation in the complex frequency domain. Expanding these effects to more complex systems necessitates new methods to study advanced scattering phenomena.

When analyzing complex structures and systems, one often brings in the concept of EM circuits grounded in circuit theoretic principles. It analyzes electric current flow and electric potential across components like resistors, inductors, and capacitors. The 'lumped element model' central to this approach simplifies each component's function, focusing on connectivity over internal mechanics, and has been particularly useful in radio-frequency and microwave applications for designing and discovering new functionalities [20]. The lumped element concept has recently been extended to higher frequencies, including terahertz, infrared, and visible light, facilitating the modeling of optical systems [21]–[25]. This method has led to discoveries and applications in these domains, enhancing the understanding of active systems [26], PT-symmetry [27]–[30], exceptional points [28], [31], topologically nontrivial states [32], [33], coherent perfect absorption [33], and BICs [34], [35]. An important question is whether this model can be adapted for the complex frequency domain.

In this work, we investigate scattering anomalies, including virtual perfect absorption, bound states in the continuum, both real and virtual critical coupling, exceptional points, and the related anisotropic transmission resonances (ATRs) within the complex frequency domain. This investigation is conducted by applying the circuit model, particularly utilizing the lumped element approach. This method significantly enhances the understanding of light scattering within the complex frequency domain. Our approach not only aids in deciphering the intricacies of these phenomena but also lays a foundational framework for developing sophisticated optical devices and systems. The versatility of this methodology is such that it can be adapted to a wide range of systems, regardless of their complexity.

**II. Virtual critical coupling and perfect absorption**

Let us examine a lossless LC resonant circuit illustrated in **Figure 1(a)**. In the lumped elements, when the elements are sufficiently small compared to the wavelengths, the circuit exhibits a single mode characterized by a real-valued (angular) eigenfrequency, $\omega_0 = 1/\sqrt{LC}$. When excited, the current and voltage oscillate at this real eigenfrequency without decay, with the electric field in the capacitor ($C$) and the magnetic field in the inductor ($L$) being phase-offset by $\pi/2$. When the circuit is connected to a transmission line, it may lose energy through leakage and reflection. The reflection coefficient is defined as $r(\omega) = \frac{Z_L(\omega) - R_0}{Z_L(\omega) + R_0}$, with $R_0$ being the characteristic impedance of the port transmission line ($R_0 = 50\ \Omega$) and $Z_L(\omega) = R_L(\omega) + jX_L(\omega)$ being the load (resonant circuit) input impedance, $j = \sqrt{-1}$, $X_L(\omega)$ is an effective frequency-dependent reactance, seen at the load terminals and $R_L$ is the load resistance.

In the purely reactive load regime ($R_L = 0$) and for the parallel connection of the inductor and capacitor, the load reactance ($X_L$) is [36] $X_L(\omega) = \omega L \frac{\omega_0^2}{\omega_0^2 - \omega^2}$. Thus, this circuit allows a fully analytical treatment with the following results for the complex reflection coefficient: $r(\omega) = -\frac{(\omega^2 - \omega_0^2) + j(\tau\omega_0^2)\omega}{(\omega^2 - \omega_0^2) - j(\tau\omega_0^2)\omega}$, where $\tau = L/R_0$. The case of the series connection of L and C differs only in the sign and value of $\tau$ [36]. Solving this equation for $r(\omega) = 0$ yields frequencies of zero reflection, $\omega_z = -j\frac{\omega_0}{2}\left(\tau\omega_0 \pm \sqrt{\tau^2\omega_0^2 - 4}\right)$. Similarly, for the pole we get, $\omega_p = j\frac{\omega_0}{2}\left(\tau\omega_0 \mp \sqrt{\tau^2\omega_0^2 - 4}\right)$. Depending upon the value of $\tau\omega_0$, we obtain three scenarios: $\tau\omega_0 > 2$, when the pole-zero pair is symmetrically distributed along the imaginary frequency axis (over coupled), $\tau\omega_0 = 2$, when the pole-zero pair coincides with the $\omega_0$ but on the imaginary frequency axis with zero real part (critically coupled), and $\tau\omega_0 < 2$, when the pole-zero pair have nonzero real and imaginary parts (under-coupled). In the latter case, the negative (positive) sign in $\omega_z$ ($\omega_p$) represents the zero (pole) in the negative half-space and will be omitted, as it is a mirror reflection of the zero (pole) with a positive-valued real part of the frequency. This latter case is of utmost importance for applications, as here, the zero and pole possess nonzero real frequencies, making them accessible to waves and signals. We will assume this scenario in the following analysis.

Now, assuming a port weakly coupled to the resonator ($\tau\omega_0 \to 0$) and calculating the reflection coefficient in the complex frequency plane reveals a pole ($|r| \to \infty$), and a corresponding complex-conjugated zero ($|r| \to 0$) located very close to the same point on the real frequency axis ($\omega_0$). In the extreme scenario where the coupling is absent, $\tau\omega_0 = 0$, the convergence of the pole and zero on the real axis occurs, leading to the formation of a bound state characterized by a diverging quality factor (Q-factor). Increasing the circuit's coupling to the port (increase of $\tau\omega_0$) causes the pole and zero to move further apart in the complex plane, **Figure 1(b)**. A deeply located zero in the complex plane implies that a circuit excited by a real frequency (CW) signal or a spectrum of signals will mostly reflect, indicating a mismatch with the port.

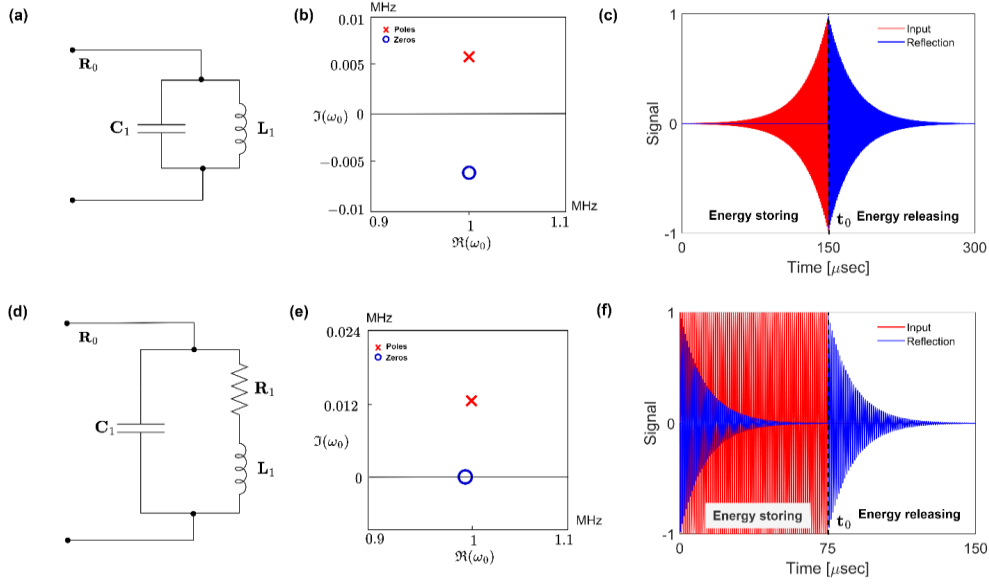

*Figure 1. Illustration of operational regimes of a resonant LC circuit. (a) Circuit for virtual perfect absorption regime (VPA) in a single port network. (b) Mismatched regime with the pole and zero located in the complex plane. (c) Transient dynamics of the resonator in the VPA regime of excitation by the complex frequency signal at the frequency of zero reflection. (d) Circuit of a critically coupled resonator with adjusted resistance ($R_1$) for minimal reflection. (e) Poles and zeros for critically coupled resonator. (f) Transient dynamics of the critically coupled resonator, highlighting the behavior of reflected signal during and after continuous-wave excitation.*

In many practical applications, including perfect absorbers [37], [38], antenna systems [39], and light coupling to ultrathin and 2D materials [40]–[42], it is essential for the circuit to match the port with minimal reflection within a certain frequency range. In one-port systems, this can be achieved only by introducing some losses through resistance ($R_1$) **Figure 1(d)**, which brings the reflection zero to the real frequency axis, **Figure 1(e)**. Upon reaching the axis, the circuit becomes *critically coupled* to the port, a state always accompanied by energy loss and referred to as "real critical coupling". This operation regime is of interest for resonant perfect absorbers [37]. For instance, **Figure 1(f)** shows the transient dynamics of such a critically coupled resonator simulated using the Advanced Design System (ADS). Here, the resonator's zero ($|r_1| = 0$) is brought to the real axis by adjusting the resistor $R_1$. The simulation shows that upon initiating continuous wave (CW) excitation, the resonator generates a reflected signal that exponentially decays as the mode builds up. After the transient regime, the reflection drops to zero. However, when the excitation is turned off at time $t_0 = 75$ μsec, a reflected signal reappears as the stored energy is released. In this work, we simulated the circuits using the PathWave ADS with its transient/convolution solver, allowing for SPICE-type transient time-domain analysis. Our analysis starts with a user-defined time-domain source, ensuring the max time step was adequate to sample the circuit's highest expected frequency. The analysis was achieved using MATLAB. The simulation employs the following parameter values: $L = 0.1$ μH, $C = 0.253$ μF, and $R_1 = 7.9$ mΩ.

Returning to the mismatched regime shown in **Figure 1(b),** where the pole and zero are situated in the complex plane, another method to achieve the zero involves using a complex frequency ($\omega = \Re(\omega) + j\Im(\omega)$), causing the excitation signal to grow exponentially, $\exp[j\omega t] = \exp[j\Re(\omega)t] \cdot \exp[-\Im(\omega)t]$. In this regime, as elaborated in optical systems like Fabry-Perot resonators [8], [11], [12], [43], the reflected signal at any given moment cancels out due to destructive interference with the incoming excitation signal. As long as this complex excitation signal is used, the system absorbs the incoming wave despite the absence of losses, **Figure 1(c).** The energy is effectively trapped within the resonator, giving rise to the *virtual perfect absorption (VPA)* effect. Once the complex frequency excitation is stopped ($t_0 = 150$ μsec), the signal becomes finite and Fourier-transformable, thus interacting with the circuit as a spectrum of real frequency signals, resulting in the emergence of a reflected signal.

It should be noted that the terminology of perfect absorption (also known as critical coupling) and virtual perfect absorption (also known as virtual critical coupling) is consistently used for the single-port circuit discussed here. In a two-port system, critical coupling refers to the absence of reflection at the circuit terminals (while allowing transmission), whereas perfect absorption implies no reflection or transmission. The concept of virtual critical coupling (virtual perfect absorption) in one circuit resonator has been experimentally investigated in [36], showcasing impedance matching in purely reactive loads through complex excitations. In another work [44], the VPA in lossless resonators was achieved by modulation of the resonator's radiative decay rate.

### III. Bound states in the continuum

Bound states in the continuum (BICs) are an extraordinary wave phenomenon that challenges conventional wisdom about wave behavior [5], [45]. Proposed by von Neumann and Wigner in 1929 [46], BICs are states that remain localized (real-valued eigenvalues) within a spectrum of radiating waves without emitting energy. The essence of BICs lies in their coexistence with extended waves while remaining perfectly confined without any radiation loss. These states arise from diverse mechanisms, including symmetry and separability in systems, parameter tuning, and inverse design methods. This phenomenon has been identified in various wave domains, including electromagnetic, acoustic, water, and elastic waves. Photonic structures are particularly important for BIC research due to their tailorable properties. BICs have significant applications in technology, particularly in creating high-performance lasers [47]–[49], sensors [50]–[53], and light state control [54]. Their unique non-radiative property has led to the development of high-Q resonators useful in optics and photonics. A BIC can be defined as a resonance with zero leakage and zero linewidth, resulting in an infinite quality factor Q. The Q-factor of a system, indicating its ability to store energy, is given by $Q = \dfrac{\Re[\omega_p]}{2\Im[\omega_p]}$, where $\omega_p$ is the complex frequency of a pole associated with the system's eigenmode. Therefore, the divergence of the Q-factor in a BIC state signifies that it originates from a pole and zero merging at the real frequency axis[8].

In a circuit model, the simplest BIC state can be realized when two resonators are coupled via a coupling capacitor and excited through a single port, as depicted in **Figure 2(a)**. The analytical analysis of this configuration using the ABCD matrices approach reveals that the system's poles and zeros exist in pairs, with the position of one pair being the complex conjugate of the other, **Figure 2(b)**. With the introduction of a small loss in the second resonator, one pair of pole and zero nearly coalesce at the real frequency axis, provided the coupling capacitance is within the microfarad ($\mu F$) range. This state, where they almost coincide on the real axis, is known as a quasi-BIC state[55]–[57], characterized by a high but finite Q-factor. In contrast, for the lossless case, achieving a true BIC state requires capacitance in the millifarad (mF) range to ensure that the poles and zeros fully coalesce on the real axis. However, a larger capacitance can be a limitation in some applications. The parameters used in **Figure 2** are $L_1 = 0.1$ µH, $L_2 = 0.5$ µH, $C_1 = C_2 = 0.253$ µF, $C_c = 10$ µF, and $R_2 = 0.0317$ mΩ.

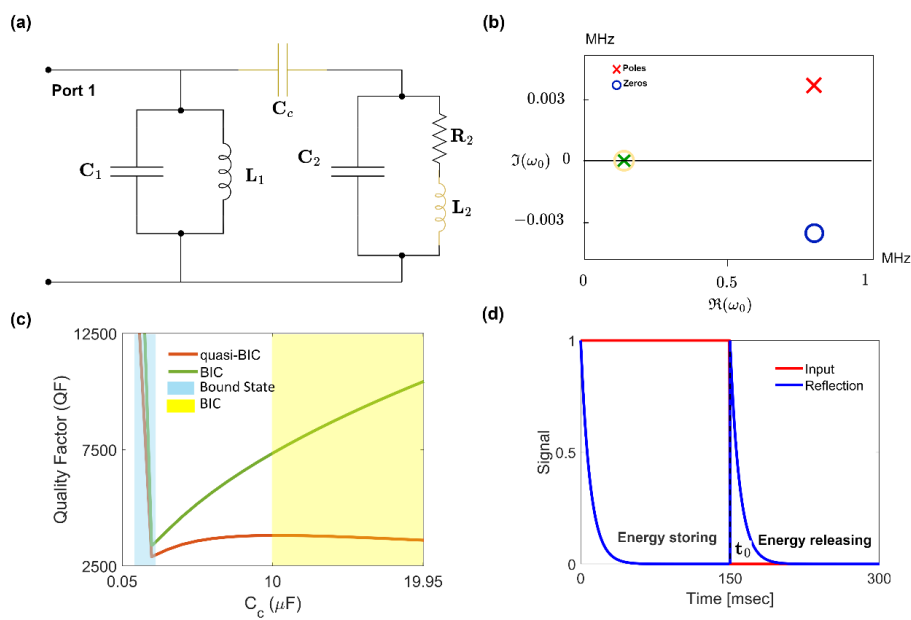

*Figure 2. Analysis of BIC in a coupled resonator system. (a) Circuit schematic for BIC realization with two resonators and a coupling capacitor, excited via a single port. (b) Poles (green) and zeros (yellow) converging on the real frequency axis in a quasi-BIC state. The poles and zeros are calculated analytically using the ABCD matrix formalism. (c) Q-factor dependence on the coupling capacitance ($C_c$) for quasi-BIC and BIC states. (d) Transient response in the coupled system, showing initial strong reflection due to resonator-port mismatch, which diminishes as the system reaches a steady state.*

**Figure 2(b)** illustrates the pole (in green) and zero (in yellow) nearly converging at the real frequency axis. The Q-factor of the coupled resonator system for both quasi-BIC and BIC cases as a function of $C_c$ is shown in **Figure 2(c)**. In the weak coupling regime (small $C_c$), the second resonator is almost uncoupled from the rest of the circuit, associated with increasing Q-factor upon reduction of $C_c$ (blue area in **Figure 2(c)**). This is an example of a discrete bound state or a state of a closed lossless resonator that, as mentioned above, is characterized by a diverging Q-factor. As the coupling capacitance $C_c$ increases, the Q-factor initially drops to its minimum value in both cases. With further increases in $C_c$, the Q-factor steadily rises for the BIC. However, in the quasi-BIC case, it initially ascends before gradually declining (around $10\ \mu F$) as the coupling intensifies. Since the Q-factor is defined by the system's poles, in the quasi-BIC case, the pole in the upper complex plane begins approaching the real axis, reducing the denominator of $r(\omega)$ and thus increasing the Q-factor. With a continued increase in $C_c$, the poles start moving to the lower complex plane, which raises the denominator of $r(\omega)$ and consequently decreases the Q-factor. The differing behaviors in these cases originate from the presence of a dissipative element ($R_2$) in the quasi-BIC scenario, in contrast to the lossless resonators in the BIC case. For a critical $C_c = 10\ \mu F$, the Q-factor for the quasi-BIC system falls within the range of $10^3$.

**Figure 2(d)** depicts the transient response of the coupled resonator system when excited with a monochromatic signal (in red), as illustrated by the envelope of the absolute-value-squared signal, which corresponds to a zero on the real axis (indicated in yellow in panel (b)). In this single-port coupled resonator system, a significant mismatch exists between the coupled resonators and the port, resulting in strong reflection, as indicated by the envelope of the absolute-value-squared signal (in blue), during the initial period. As the modes within the system build-up, the reflection steadily decreases until it diminishes to zero. Being a one-port network, the energy is stored in the resonators. Immediately after the cessation of the signal at $t_0$, the condition for zero reflection is no longer met, leading to the release of the energy stored in the resonator. Nevertheless, due to our operation within the quasi-BIC regime, the attenuation of this quasi-bounded mode is significantly slower compared to a single resonator, with the decay timescale extending to milliseconds as opposed to microseconds in Figure 1.

### IV. Critical coupling in two-port systems

Transitioning to systems with more than one port, the scattering matrix expands to a 2x2, $\hat{S} = \begin{bmatrix} S_{11} & S_{12} \\ S_{21} & S_{22} \end{bmatrix}$, which equals $\hat{S} = \begin{bmatrix} r_1 & t \\ t & r_2 \end{bmatrix}$ in magnet-less reciprocal systems, like in this work. Here, $r_1$ and $r_2$ represent reflection coefficient from ports 1 and 2, respectively, and $t$ is the transmission coefficient, identical for both directions due to reciprocity [58]. In such 2-port systems, it is important to distinguish between zeros of reflection coefficients ($|r_{1,2}| = 0$) and zeros of the entire scattering matrix $\hat{S}$. For the latter, zeros can be identified, for example, as points where the determinant of the matrix is zero, i.e., $\det(\hat{S}) = 0$. However, the poles, which represent the system's eigenmodes, are identical in both cases [8].

The zeros in the reflection and scattering matrix signify two distinct effects. A zero in the reflection coefficient indicates a no-reflection regime at the corresponding port. This can occur in circuits even without dissipative elements like resistors, as a second port allows scattering into this new channel, bringing the zeros to the real frequency axis. In other words, the second port produces a reflected wave that destructively interferes with the wave reflected from the first port, facilitating perfect transmission. The spectral bandwidth of such a circuit, functioning as an impedance transformer, is determined by the resonant characteristics of its elements.

In the case of scattering matrix zeros, characterized by $\det(\hat{S}) = 0$, which is achieved by exciting the system with a specific state vector—implying simultaneous excitation from both ports with amplitude- and phase-tailored pulses—the zeros correspond to the *coherent perfect absorption* (CPA) regime [8], [37], [59]–[61]. A CPA generalizes perfect absorbers to waveforms consisting of two or more waves, such as that incident on opposite faces of an open slab, film, or, as in our case, a 2-port circuit. Achieving perfect absorption requires carefully selecting system parameters, operating wavelength, and input waveform, including the intensities and relative phases of the input signals. CPAs' sensitivity to the input waveform offers flexible control over light scattering and absorption, a phenomenon demonstrated in RF, microwave [62], acoustic [63], quantum optics [64], and optical domains [37].

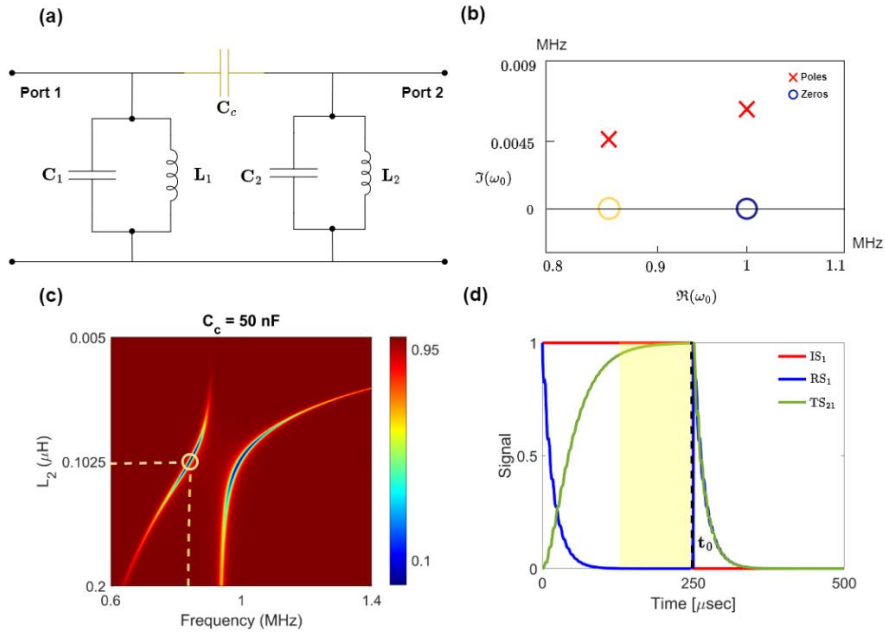

*Figure 3:* (a) Schematic of a dissipation-less 2-port circuit designed for critical coupling, comprising two resonators connected via a coupling capacitor $C_c$. (b) Distribution of poles (red) and zeros (yellow and blue) in the complex frequency plane, highlighting the impact of mode hybridization. The poles and zeros are calculated analytically using the ABCD matrix formalism. (c) Reflectance as a function of real frequency and inductance ($L_2$) for a fixed coupling capacitance of 50 nF. (d) Transient response (absolute-valued-squared) of the coupled LC resonator circuit when excited at a zero (yellow) on the real frequency axis.

Consider the dissipation-less circuit shown in **Figure 3(a)**, a 2-port system comprising two resonators, similar to those discussed earlier, connected via a coupling capacitor $C_c$. In this resistor-free circuit, energy dissipation occurs only in the form of radiation from the ports, namely reflection and transmission. Consequently, the scattering matrix zeros $\det(\hat{S}) = 0$ are situated in the complex frequency plane, specifically in the lower complex plane. However, the circuit can be designed to bring the reflection zeros ($|r_{1,2}| = 0$) to the real frequency axis through the previously discussed mechanism of destructive interference. The parameters of the designed circuit are: $L_1 = L_2 = 0.1$ $\mu$H, $C_1 = C_2 = 0.253$ $\mu$F and $C_c = 50$ nF. **Figure 3(b)** displays the system's poles and reflection coefficient zeros calculated analytically using the ABCD matrix formalism. Both zeros are located at the real axis due to the circuit's symmetry. The coupling between the resonators' modes, facilitated by $C_c$, leads to mode hybridization with characteristic repulsion of the poles and corresponding zeros in the complex plane. In this setup, coupling makes identical resonators with equal eigenmodes acquire different real frequencies. This mode hybridization is manifested as an avoided crossing in the reflection spectrum at real frequency excitation, shown in **Figure 3(c)**, indicative of the *strong-coupling regime* [65] where mode coupling surpasses dissipation, leading to mode hybridization through rapid energy exchange between resonators [66], [67]. Varying a parameter in this circuit while keeping $C_c$ fixed offers further insights. The density plot in **Figure 3(c)** maps the reflectance as a function of real frequency and inductance $L_2$. Altering this parameter results in spectral detuning of the modes and changes in coupling strength due to varying spectral overlap.

When excited by a continuous wave (CW) at these real frequency axis zeros, the transient response of the coupled resonator system is observed. **Figure 3(d)** presents the transient response when excited at one of these zeros (yellow in panel (b)), focusing only on the signal envelopes. After a brief transient period, the system stabilizes, characterized by the absence of scattering, signifying complete energy transmission through the circuit. Upon cessation of excitation (at $t_0 = 250$ μsec), the energy used to build up the modes is released.

## V. Virtual PT-symmetry and Exceptional Points

The concept of PT-symmetry, originally stemming from quantum mechanics, holds significant implications in electrical engineering[68]. Traditional quantum mechanics, under the Dirac-von Neumann formulation, relies on Hermitian operators for describing systems that correlate with observable, real-valued physical phenomena. However, this perspective has evolved with recognizing non-Hermitian systems in optics and electromagnetics, where phenomena such as material loss and gain or radiation loss in open resonators are prevalent. The pioneering work by Bender and Boettcher demonstrated that Hermiticity is not essential for obtaining real eigenvalue spectra, thereby bridging a connection between quantum mechanics and electromagnetism through the concept of PT-symmetry [69]–[72]. The application of PT-symmetry has been explored in various aspects of electrical engineering, including optical waveguides, coupled resonators, photonic crystals, and metamaterials [73]. These implementations have broadened the scope of PT-symmetry, extending its influence to areas like atomic systems, electronics, and acoustics.

The role of Exceptional Points (EPs) [74] in PT symmetry is of particular interest in electrical engineering [7], [73], [75], [76]. EPs signify a phase transition where eigenvalues and corresponding poles shift from real axes to complex plane, revealing the non-Hermitian nature of the systems. This transition at EPs leads to a plethora of novel applications and phenomena in electrical engineering, such as enhanced sensing capabilities, development of exotic lasing techniques, advancements in optical nonreciprocity and isolation, loss-induced transparency, and wireless power transfer [68], [73], [74], [77].

Photonic PT-symmetric structures, which maintain invariance under both time and parity inversion, are achieved by meticulously balancing loss and gain within the system [7], [8], [75]–[77]. An exemplary instance of a PT-symmetric system can be observed in a coupler that consists of two interconnected components, such as optical waveguides/cavities. These elements are identical except that one element possesses gain while the other shows an equivalent amount of loss. Let's consider a two-resonator system with different resonant frequencies $(\omega_{0,1}, \omega_{0,2})$ and a decay rate $\gamma_1, \gamma_2$ associated with both resonators. The dynamics of such a coupled system is described by [78]

$$\begin{pmatrix} \dot{a}_1 \\ \dot{a}_2 \end{pmatrix} = -j \begin{pmatrix} \omega_{0,1} - j\gamma_1 & \kappa_{12} \\ \kappa_{21} & \omega_{0,2} - j\gamma_2 \end{pmatrix} \begin{pmatrix} a_1 \\ a_2 \end{pmatrix} \quad (1)$$

where $a_1$ and $a_2$ are the localized resonator modes amplitudes. The parameters $\kappa_{12}$, $\kappa_{21}$ describe the coupling strength between the two resonators and $\gamma_i$ accounts for the loss for each resonator modes. The energy conservation imposes a restriction on $\kappa_{12}$ and $\kappa_{21}$ such that for the lossless case: $\kappa_{12} = \kappa_{21}^*$ [78], [79]. We set $\kappa_{12,21} = \kappa$ with $\kappa \in \mathbb{R}$ to justify this requirement. We also assume that the intrinsic eigenfrequencies of both modes are equal, $\omega_{0,1} = \omega_{0,2} = \omega_0$. Then the effective Hamiltonian of the system is $\hat{H}_0 = \begin{pmatrix} \omega_0 - j\gamma_1 & \kappa \\ \kappa & \omega_0 - j\gamma_2 \end{pmatrix}$. This Hamiltonian is PT-symmetric, $(\hat{P}\hat{T})\hat{H}_0 = \hat{H}_0$, if $\gamma_1 = -\gamma_2$. The standard procedure of finding eigenvalues of the Hamiltonian thereby yielding [68], $\omega_\pm = \omega_0 - j\gamma_{ave} \pm \sqrt{\kappa^2 - \gamma_{dif}^2}$, where $\gamma_{ave} = (\gamma_1 + \gamma_2)/2$, and $\gamma_{dif} = (\gamma_1 - \gamma_2)/2$. In particular, when $\gamma_1 = -\gamma_2 = \gamma$, the eigenvalues of the system are: $\omega_{PT} = \omega_0 \pm \sqrt{\kappa^2 - \gamma^2}$. If the coupling in the system is sufficiently strong, that is, when $\kappa > \gamma$, the eigenvalues are purely real, giving rise to the PT-symmetric phase. If the opposite case of $\kappa < \gamma$, the decay overcomes energy exchange between the modes, the system is called to be in the PT-symmetry broken phase. The balanced phase, $\kappa = \gamma$, corresponds to the EP [8], [74], [76], [77].

Achieving the equilibrium necessary for PT-symmetry and EP is a complex task due to the difficulty in maintaining a balanced distribution of loss and gain, which becomes increasingly challenging at higher operational frequencies. However, certain PT-symmetry phenomena can still be investigated in non-Hermitian systems with an offset in loss and structures that only incorporate loss. The Hamiltonian derived above for a general, lossy, two-mode system, $\hat{H}_0$, can be decomposed into a PT-symmetric part ($\hat{H}_{PT}$) and a decaying part ($\hat{H}_L$)[68]:

$$\hat{H} = \begin{pmatrix} \omega_{0,1} - j\gamma_{dif} & \kappa \\ \kappa & \omega_{0,2} + j\gamma_{dif} \end{pmatrix} - j\begin{pmatrix} \gamma_{ave} & 0 \\ 0 & \gamma_{ave} \end{pmatrix}.$$ Here, $\hat{H}_{PT}$ is analogous to the PT-symmetric non-Hermitian Hamiltonian for balanced gain and loss rates. This formulation enables the observation of PT symmetry effects and EPs in passive systems with a loss offset, simplifying the creation of PT-symmetric systems [71], [80]–[83]. The EP in such passive analogs of PT-symmetry is located in the upper complex half-plane and can be detected through parameter variations or complex excitations [13]. In experiments with loss-only structures, a phase transition is observed as a change in mode symmetry, shifting from asymmetric in the broken symmetry regime to symmetric in the PT-symmetric phase when material loss is adjusted [71]. However, it is important to note that systems with only loss and those with balanced gain-loss in PT symmetry differ significantly. For instance, in solely lossy systems, phenomena like lasing or lasing-CPA, common in PT-symmetric systems [8], [75], are unattainable. Moreover, EPs can arise in systems devoid of loss and gain, which are inherently Hermitian [84], [85].

Now, we consider the design of a virtual EP regime in a purely lossy electric circuit, as depicted in **Figure 4(a)**. This circuit comprises two resonant contours, which provide the necessary modes for the EP state. The first resonator is entirely lossless, possessing an eigenfrequency $\omega_{0,1} = 1/\sqrt{C_1 L_1}$. In contrast, the second resonant contour includes a resistive element $R_2$, resulting in a complex eigenfrequency. Applying Kirchhoff's voltage law yields the following complex eigenfrequency for the lossy resonator: $\omega_{0,2} = \frac{1}{2}\left( \frac{\sqrt{4L_2 - C_2 R_2^2}}{L_2 \sqrt{C_2}} + j\frac{R_2}{L_2} \right)$. Expanding the expression using the Taylor series, we can approximate the eigenfrequency as $\omega_{0,1} = \frac{1}{\sqrt{C_2 L_2}} + j\frac{R_2}{2L_2}$. Consequently, due to losses, the eigenfrequency and the corresponding pole shift to the upper complex plane by a value of $\frac{R_2}{2L_2}$.

The resonators are interconnected via the coupling capacitor $C_c$. The coupling strength of the system, is described using coupled-mode theory, is given by $\kappa = \frac{\sqrt{\omega_{0,1} \omega_{0,2}}}{\sqrt{C_1 C_2}} \frac{C_c}{2}$ [79]. Therefore, by assessing the coupling strength $\kappa$ and decay rate, we can deduce that $\kappa \cong \gamma$ for $C_c = 2.2$ nF (near EP regime), and $\kappa > \gamma$ for $C_c = 20$ nF (strong coupling regime) in our case.

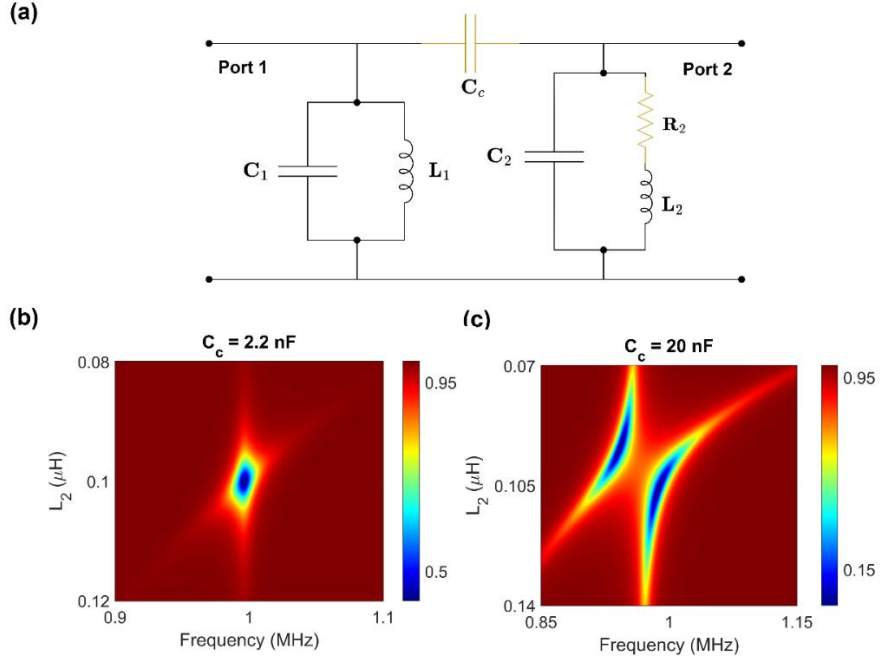

*Figure 4.* (a) Circuit schematic of a two-port coupled resonator system designed for realizing a virtual exceptional point. (b) Density plot showcasing the reflectance as a function of frequency and inductance ($L_2$) in the regime $\kappa \cong \gamma$ for $C_c = 2.2$ nF (near EP). (c) Density plot illustrating the reflectance as a function of frequency and inductance in the regime $\kappa > \gamma$ for $C_c = 20$ nF (strong coupling regime or unbroken passive PT-symmetry).

The density plots in **Figures 4(b)** and **(c)** display the reflectance as a function of real frequency and inductance ($L_2$) in the two characteristic regimes of coupling $\kappa \cong \gamma$ ($C_c = 2.2$ nF) and $\kappa > \gamma$ ($C_c = 20$ nF), respectively. With a coupling capacitance of 2.2 nF, the system exhibits a single resonant mode, indicative of, as we will see below, the EP where both poles coalesce in the complex plane. As the coupling between the resonators increases, an avoided crossing is observed between the two modes, as seen in **Figure 4(c)**, which is characteristic of strong coupling in the system. In this regime, the system experiences fast energy exchange between the two modes and their hybridization. The simulations use characteristic parameter values of $L_1 = L_2 = 0.1$ µH, $C_1 = C_2 = 0.253$ µF, $C_c = 2.2$ nF and $C_c = 20$ nF, and $R_2 = 0.01$ Ω.

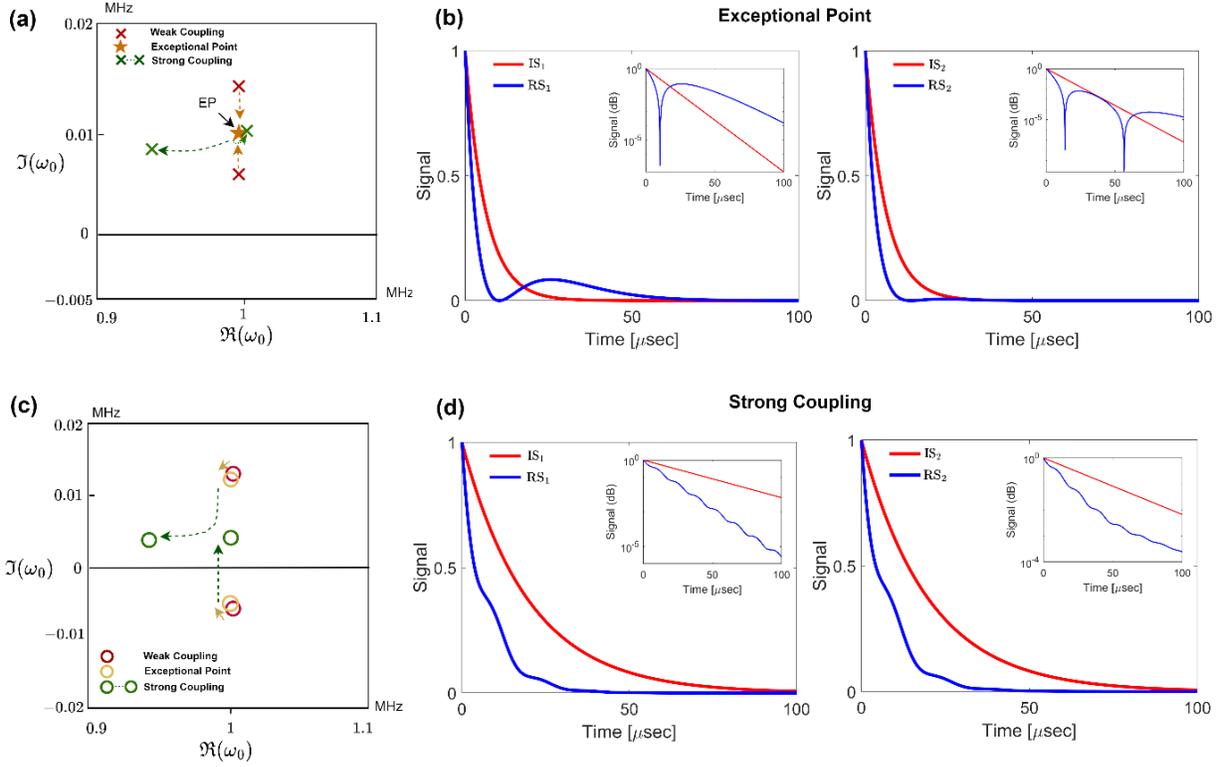

*Figure 5. (a) Illustration of pole locations in the complex-frequency plane, characterizing different coupling regimes: weak coupling, strong coupling, and EP. (b) Incident (in red) and reflected (in blue) signals (absolute-value-squared) for the EP coupling regime. (c) Display of zero locations in the complex-frequency plane, providing insights into the system's response under varying coupling conditions. The poles and zeros are calculated analytically using the ABCD matrix formalism. (d) Incident (red) and reflected (blue) signals (absolute-value-squared) in the strong coupling regime, demonstrating symmetrical and minimized reflection from both ports in the quasistationary state. The inserts show the signals in the log scale.*

The complex frequency plane's poles and zeros for the coupled resonator circuit are depicted in **Figures 5(a)** and **(c)**, marking the characterization of three distinct coupling regimes: weak (red), near-EP (yellow), and strong (green). **Figures 5(b)** and **(d)** illustrate the system's behavior when signals are incident from both port 1 and port 2 in different coupling regimes. $IS_1$ ($IS_2$) and $RS_1$ ($RS_2$) denote the incident signal and reflection signal from the first (second) port. In **Figure 5(b)**, under the weak coupling regime, the reflected signal from both ports initially decays to zero. However, as the input signal stabilizes in the quasistationary regime, the reflected signal from port 1 increases while vanishing in port 2. **Figure 5(d)** showcases the strong coupling regime, where the reflected signal from both ports rapidly decays to zero after a brief transient period and remains at zero in the quasistationary state. This behavior signifies an energy exchange between the resonators and consistent reflection characteristics, regardless of the port receiving the incident signal. In this regime, the strong coupling leads to a symmetrical response in both ports.

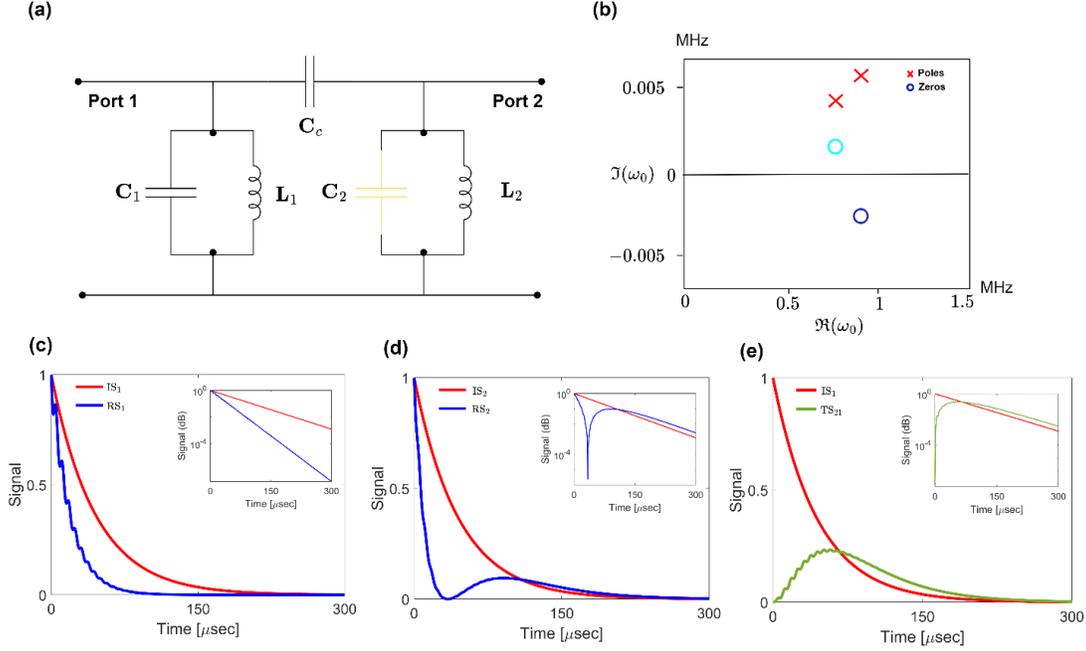

*Figure 6. (a) Circuit schematic employed to demonstrate the virtual ATR effect. (b) Location of the zero in the upper half of the complex-frequency plane, crucial for ATR, shown in sky blue. (c) Reflected signal $RS_1$ (absolute-value-squared) decaying to zero in the quasistationary state, illustrating the ATR effect when excited from one port. (d) Reflected signal $RS_2$, depicting a rapid decay followed by an increase and eventual exceedance of the input signal in the quasistationary state when excited from the opposite port. (e) Transmitted signal $TS_{21}$, highlighting the "virtual gain" effect where the transmitted signal surpasses the excitation in the quasistationary regime. The inserts show the signals in the log scale.*

## VI. Virtual Anisotropic Transmission Resonance and Virtual Gain

The coalescence of poles in the complex plane, although difficult to detect directly by an observer in experiments, can be inferred by observing related effects accompanying the EP regime. In this work, we investigate the anisotropic transmission resonance (ATR)[13], [86], characterized as unidirectional invisibility in one-dimensional systems. To achieve this regime, the system under consideration must exhibit vanishing reflectance from one port and very strong reflectance from the opposite port, accompanied by a high transmission that remains identical for both directions in reciprocal systems. In other words, ATR is a special type of scattering process that conserves flux for wave incident from only one port, which occurs when either of the reflectance's becomes zero, and transmission turns to one, following the conservation relation $|T - 1| = \sqrt{\mathcal{R}_1 \mathcal{R}_2}$, where $T = |t|^2$ is the transmittance and $\mathcal{R}_i = |r_i|^2$ is the reflectance in the port $i$ [86].

The conservation relation does not apply to complex frequency excitation, i.e., for virtual ATRs [13]. The virtual ATR effect has been theoretically predicted for passive PT-symmetric structures [13]. This state can be achieved by exciting the system with a corresponding complex frequency. In the virtual ATR regime, the transient reflected signal from one port of a 2-port system quickly drops to zero upon excitation by a complex decaying signal. In contrast, reflectance from the opposite port increases and may even surpass the transmitted signal. Here, we demonstrate that by meticulously designing the circuit depicted in **Figure 6(a),** achieving the conditions for virtual ATRs is feasible without the need for a balanced gain and loss in the system. The circuit comprises two slightly detuned, lossless resonators weakly coupled through a coupling capacitance. The simulation utilizes characteristic parameters with $L_1 = L_2 = 0.1$ µH, $C_1 = 0.253$ µF, $C_2 = 0.3$ µF, and $C_c = 50$ nF. It is crucial to note that the phenomenon of ATR can be observed within a passive PT symmetric system in the EP regime, where the reflection vanishes from one side depending upon the port of excitation, as illustrated in **Figure 5(b)**. However, due to the losses in the circuit, the observed effect is not pronounced. To realize ATR in a two-port network, the circuit is modified with lossless resonators instead of lossy, with a slight alteration in their resonant frequencies. Such a modification optimizes the circuit for an effective realization of ATR.

To demonstrate the ATR effect in the circuit shown in **Figure 6(a),** we excite the circuit at its zero, located in the upper half of the complex-frequency plane (depicted in sky blue in **Figure 6(b)**). The incident signal is tailored to match this zero, i.e., a complex frequency excitation with an exponentially decaying envelope (illustrated in red), as seen in **Figures 6(c)-(e). Figure 6(c)** demonstrates that $RS_1$ decays and ultimately diminishes to zero in the quasistationary state. Conversely, when excited from the other port, $RS_2$ (**Figure 6(d)**) initially decays rapidly to a minimum value following a short transient period, then increases, and in the quasistationary state, it follows and even surpasses the input signal. **Figure 6(e)** displays the transmitted signal, equal for both excitation cases. The transmitted signal also exceeds the excitation in the quasistationary regime, showcasing an effect now termed "*virtual gain*." [13], [14], [17], [19]. This phenomenon, where the scattered field (reflection and transmission, in our 1D system case) exceeds the excitation, akin to active media with material gain, is attributed to the initial storage and subsequent release of energy. Thus, the scattered signal becomes more significant than the exponentially decaying excitation. Consequently, we observe an ATR characterized by a response in which reflection from one port is null while yielding significant reflection from the other port.

## VII. Multi-Resonator System

Finally, to explore more intricate poles and zeros configurations, we apply the approach to a more complex system incorporating multiple cascaded lossy resonators, as illustrated in **Figure 7(a)**. The poles and zeros of this circuit are presented in **Figure 7(b)**. The simulation utilizes characteristic parameters with $L_1 = L_2 = L_3 = 0.1$ µH, $C_1 = C_2 = C_3 = 0.253$ µF, $R_1 = R_2 = 0.01 \,\Omega$, $R_3 = 1 \, m\Omega$ and $C_{c_1} = C_{c_2} = 10$ nF. In the case of negligible coupling between resonators, the circuit's behavior is predominantly influenced by the first resonator's poles and zeros. As the coupling intensifies, it causes the poles and zeros to move further apart within the complex frequency plane. **Figures**

**7(c)-(e)** showcase the virtual perfect absorption regime where the system is excited by exponentially decaying signal aligned with the corresponding zeros. The reflection coefficient ($r_1$) decays to zero as the system stabilizes. Note that the reflection coefficient in **Figure 7(c)** decays slower than in other cases due to the close location of the corresponding pole and zero.

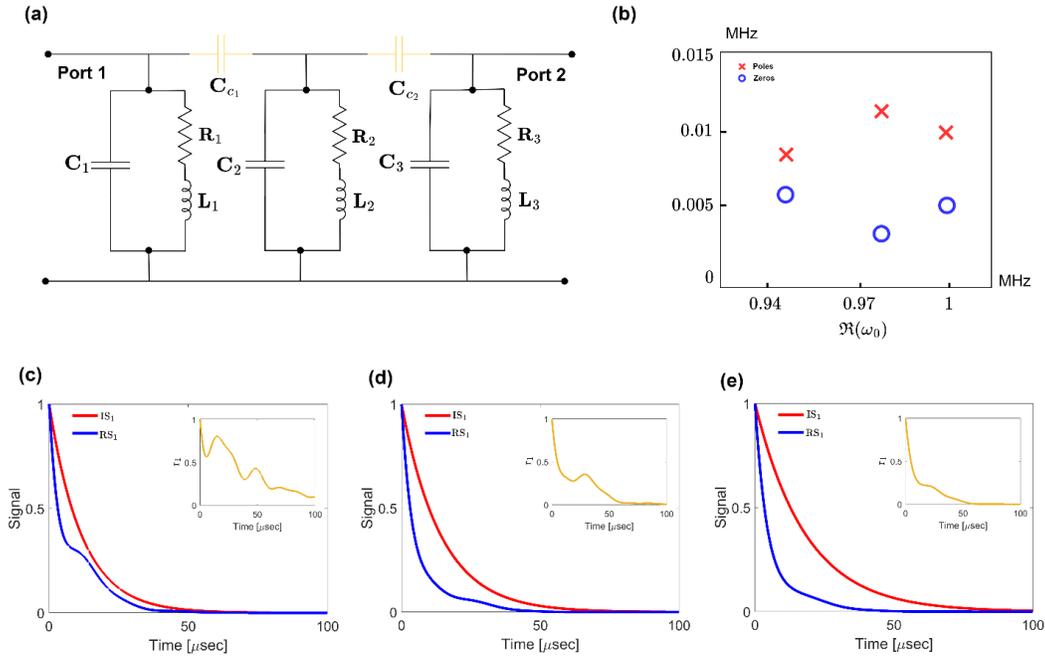

*Figure 7. (a) Schematic of the circuit employed to demonstrate VPA phenomena in a multi-resonator system. (b) Location of the poles and zeros of this circuit. (c)-(e) Incident (in red) and reflected (in blue) signal (absolute-value-squared), along with the reflection coefficient, demonstrating the VPA regime. The inserts demonstrate the reflectance ($r_1$=RS/IS) signals.*

## VIII. Conclusions

In this work, we have embarked on an in-depth exploration of EM wave scattering phenomena in the complex frequency domain, employing the principles of electrical circuits to elucidate the behavior of light scattering and the underlying physics. Our study began with a foundational examination of virtual perfect absorption, where we applied the lumped element model to understand and predict the behavior of LC circuits in various operational regimes. The zero-reflection states and the critical coupling conditions revealed in this analysis are instrumental in designing and implementing efficient devices, especially in applications demanding minimal energy loss and maximal energy utilization. Further, we delved into the intriguing concept of BICs. By employing a coupled resonator system, we demonstrated the realization of quasi-BIC and BIC states, showcasing their enormous potential in creating high-Q resonators for optical and photonic applications. Our analysis provided valuable insights into manipulating these states through coupling parameters, offering a pathway to harnessing their unique non-radiative properties for technological advancements. Transitioning to two-port systems, we examined the critical coupling phenomenon and its implications for wave scattering and absorption. Our study highlighted the distinct effects of zeros in reflection coefficients and scattering matrices, leading to novel applications like coherent perfect absorption. The versatility of these systems in controlling light scattering and absorption at the subwavelength scale opens exciting possibilities in photonics. Our investigation then focused on PT symmetry and EPs. We demonstrated the design and realization of a virtual EP regime, paving the way for novel applications in passive photonic systems. We explored the phenomenon of virtual ATR and virtual gain. We unveiled possibilities for controlling wave propagation in photonic structures by achieving unidirectional invisibility in a two-port system. The significance of these effects in the realm of electric circuits is profound. Finally, to explore more intricate poles and zeros configurations, we applied the approach to a more complex system incorporating multiple cascaded lossy resonators. Our findings enhance understanding of complex wave phenomena and provide a versatile toolkit for designing and optimizing advanced optical devices and systems. This work is a testament to the power of interdisciplinary research, bridging the gap between electrical engineering principles and nanophotonics, enriching both fields with new perspectives and capabilities.

**Acknowledgments**

The authors thank the ECE department of Florida International University.

**Conflict of Interest**

The authors have no conflicts to disclose.